# Advanced Prediction of Hypersonic Missile Trajectories with CNN-LSTM-GRU Architectures

1st Amir Hossein Baradaran
*School of Electrical Engineering*
*Iran University of Science and Technology*
Tehran, Iran
A_Baradaran@cmps2.iust.ac.ir

*Abstract*— Advancements in the defense industry are paramount for ensuring the safety and security of nations, providing robust protection against emerging threats. Among these threats, hypersonic missiles pose a significant challenge due to their extreme speeds and maneuverability, making accurate trajectory prediction a critical necessity for effective countermeasures. This paper addresses this challenge by employing a novel hybrid deep learning approach, integrating Convolutional Neural Networks (CNNs), Long Short-Term Memory (LSTM) networks, and Gated Recurrent Units (GRUs). By leveraging the strengths of these architectures, the proposed method successfully predicts the complex trajectories of hypersonic missiles with high accuracy, offering a significant contribution to defense strategies and missile interception technologies. This research demonstrates the potential of advanced machine learning techniques in enhancing the predictive capabilities of defense systems.

## I. INTRODUCTION

Hypersonic missiles revolutionize modern warfare with their unmatched speed, agility, and precision. Traveling at speeds exceeding Mach 5, they can strike distant targets in minutes, evading traditional defense systems. Their ability to maneuver mid-flight makes interception challenging, posing significant threats to strategic assets. As a result, they are reshaping global defense strategies and prompting advancements in countermeasure technologies [1].

The AGM-183A Air-launched Rapid Response Weapon (ARRW) is a hypersonic missile developed by Lockheed Martin for the United States Air Force. Engineered to engage high-value and time-sensitive targets, the ARRW employs a boost-glide mechanism. After being launched from an aircraft, a rocket booster accelerates the missile to speeds exceeding Mach 5, at which point the glide vehicle separates and autonomously maneuvers toward its target [2]. This advanced system provides the U.S. military with a rapid-response capability, enabling stand-off attacks against heavily defended or quickly moving targets. As shown in Figure 1, the AGM-183A features a sleek design optimized for high speed and maneuverability, making it exceptionally challenging for adversaries to detect or intercept. Since November 2024, the ARRW has undergone extensive testing, including a significant trial conducted in December 2022, launched from a B-52H Stratofortress off the coast of Southern California.

The Fattah hypersonic missile, unveiled by Iran in June 2023, marks a substantial advancement in the nation's missile technology, as illustrated in Figure 2. The missile boasts a reported range of 1,400 kilometers and achieves speeds ranging from Mach 13 to 15.

Equipped with a solid-fuel engine, the Fattah missile is designed for exceptional maneuverability, enabling it to evade even the most advanced missile defense systems. Its precision guidance system enhances its capability to strike high-value targets effectively, posing a significant challenge to existing defense infrastructures [3]. In November 2023, Iran introduced an upgraded variant, the Fattah-2, featuring enhanced maneuverability and guidance technologies. The Fattah missile was deployed operationally during Operation True Promise 2 in October 2024, during which approximately 180 missiles were launched toward Israel. This operational use highlights its strategic role in modern missile warfare.

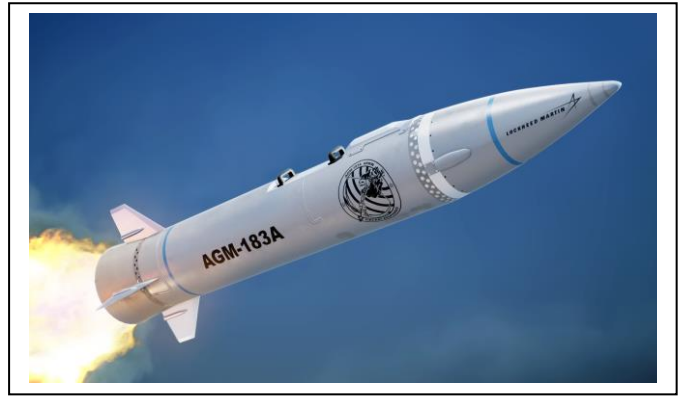

Fig. 1. AGM-183A hypersonic missile

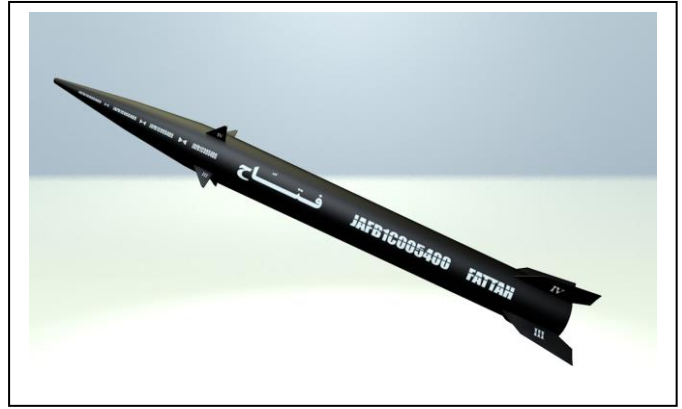

Fig. 2. Fattah hypersonic missile

The utilization of hypersonic missiles in modern warfare underscores their strategic significance and the pressing need to address the challenges they pose. Consequently, accurately predicting the trajectories of these missiles is of paramount importance to developing effective countermeasures, thereby ensuring global security and safeguarding humanity. In response to this critical need, this research introduces an advanced hybrid architecture that integrates Convolutional Neural Networks (CNN), Long Short-Term Memory (LSTM) networks, and Gated Recurrent Units (GRU). This innovative approach achieves highly accurate trajectory predictions with minimal error, offering a significant contribution to defense technologies and security systems.

Figure 3 presents a detailed internal schematic of a hypersonic missile with a scramjet engine, offering insights into the complex design and advanced technology used in modern hypersonic weapons. This figure highlights several key components that work together to enable the missile to reach speeds exceeding Mach 5 while maintaining precision, maneuverability, and stability.

At the front of the missile, the warhead is typically located, designed to carry the missile's payload, which could include conventional explosives or nuclear materials depending on the mission. This section is crucial for delivering impact on high-value targets, and its design must be optimized for minimal drag and maximum damage upon detonation.

Just behind the warhead is the guidance system, which includes navigation, control, and avionics that direct the missile along its flight path. This system enables the missile to perform evasive maneuvers, maintain course stability, and adjust its trajectory based on real-time data received from onboard sensors, such as Global Positioning System (GPS), Inertial Measurement Unit (IMU), and targeting radars. The guidance system ensures that the missile can accurately reach its target, even when flying at extreme altitudes and speeds.

The fuel tanks are situated towards the middle of the missile's body, housing the propellant necessary to sustain the missile's speed. Fuel efficiency is essential for hypersonic flight, and the advanced propulsion system must manage the flow of fuel effectively to maintain long-range performance at high speeds. The fuel system works in conjunction with the scramjet engine, which operates efficiently at hypersonic speeds by using air compression within the engine to maintain combustion at extremely high velocities. The engine's fuel subsystem ensures that the necessary energy is supplied to keep the missile in motion.

The scramjet engine is the heart of the missile's propulsion, allowing it to operate within the hypersonic regime. As a supersonic combustion ramjet, it allows for continuous high-speed flight by compressing incoming air at hypersonic speeds, igniting the fuel mixture, and propelling the missile forward. The engine subsystems in this section are responsible for maintaining optimal air pressure, combustion conditions, and fuel flow. These components are engineered to handle extreme temperatures and dynamic pressures, ensuring that the engine operates effectively throughout the missile's flight [4].

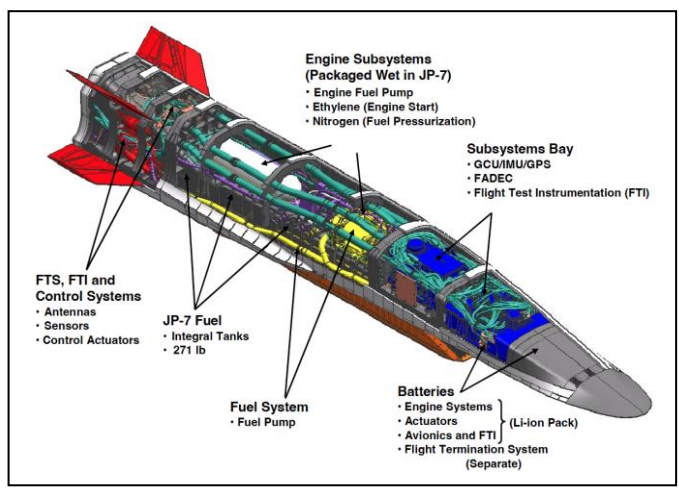

Fig. 3.  Internal Schematic of a Hypersonic Missile with Scramjet Engine

Finally, the missile's internal avionics, flight termination systems (FTS), and power supply (usually powered by batteries and fuel cells) are crucial for its navigation and control. These systems monitor the missile's performance, provide feedback to the guidance system, and ensure that the missile can be terminated safely if necessary. The flight control actuators adjust the missile's aerodynamic surfaces (e.g., fins and canards) to maintain course correction, while the FTS serves as a failsafe mechanism for missile deactivation if the missile strays off course.

In conclusion, Figure 3 provides a comprehensive breakdown of the internal structure of a hypersonic missile with a scramjet engine, showcasing the integration of advanced propulsion, control systems, and payload delivery mechanisms. The design of these systems allows the missile to maintain stable, high-speed flight and execute dynamic maneuvers, making it a powerful and difficult-to-counter weapon in modern military arsenals.

## II. RELATED WORK

Bartusiak, Nguyen, Chan, Comer, and Delp, in their paper, explore machine learning methods for the automatic identification of various hypersonic glide vehicles and ballistic reentry vehicles (RVs) using trajectory segments. Their methods, trained on aerodynamic state estimates, analyze critical vehicle maneuvers to classify vehicles with high accuracy. Additionally, they report improved identification accuracy as the time after liftoff (TALO) increases and more data becomes available for analysis. However, the reliance on an increasing amount of post-liftoff data presents a key limitation. This approach diminishes the method's utility in situations where early and precise information is critical, particularly for defensive or preemptive measures. Furthermore, their focus on classification rather than trajectory prediction leaves a significant gap in addressing scenarios where understanding the entire flight path of a missile is crucial for developing effective countermeasures. These shortcomings emphasize the importance of advancing research in missile trajectory prediction, which can provide more actionable intelligence earlier in the missile's flight [5].

Liu, Lu, Zhang, Xie, and Chen propose a novel trajectory forecasting method based on the Sparse Association Structure Model (SASM) to address the challenges in hypersonic glide vehicle (HGV) trajectory prediction. The SASM identifies relationships among known data points, transfers these associative relationships to unknown data, and enhances the model's generalization capabilities. The methodology begins by establishing a trajectory database that incorporates different maneuvering modes and models, specifically focusing on attack and bank angles of HGVs. Three trajectory parameters are then chosen as prediction variables based on the maneuvering characteristics of HGVs. Using these parameters, a prediction model grounded in the SASM framework is developed to forecast trajectory parameters. The SASM model demonstrates strong accuracy and generalization across trajectories for three distinct HGV types [6].

Yuanjie, Chunqiao, Jin, Xiaomeng, and Tianxing propose an intelligent neural network prediction model based on LSTM to predict the slippage leap trajectory of hypersonic vehicles. The study achieves high accuracy in trajectory prediction by constructing a target trajectory dataset using a high-precision hypersonic simulation model. Experimental results demonstrate that the LSTM network developed in this work can effectively predict the slippage leap trajectory of hypersonic vehicles with acceptable error margins, offering valuable insights for the development of anti-missile interception systems [7].

Despite the promising results in [6] and [7], the reliance on a single model in both studies may limit their ability to fully capture the complexity and variability of hypersonic vehicle trajectories.

A single model approach lacks the adaptability required to handle diverse and dynamic scenarios effectively. The absence of a hybrid model or a more comprehensive ensemble method potentially reduces accuracy and robustness in predicting trajectories under complex and unpredictable conditions. Therefore, adopting a hybrid approach may provide a stronger capability to reduce errors and improve predictive performance.

## III. METHODOLOGY

The primary objective of this paper is to introduce a robust model for predicting the trajectory of hypersonic missiles. Accurate trajectory prediction is critical for enhancing defensive capabilities and countermeasures against these advanced weapons. To achieve this goal, the trajectory of a hypersonic missile is first simulated, generating a comprehensive dataset containing detailed trajectory information. This dataset is meticulously designed to provide the necessary input for training and evaluating the proposed model, ensuring its performance is assessed based on specific criteria. This section offers a detailed overview of the trajectory simulation process, including the methodology used to replicate the dynamic flight behavior of hypersonic missiles. Following this, the proposed hybrid model is presented, highlighting its architecture and the integration of techniques tailored to effectively predict missile trajectories with high accuracy.

### A. Hypersonic missile trajectory

In this study, the trajectory of a hypersonic missile is simulated using a numerical integration approach to model its motion under the influence of various forces, including gravity, drag, and lift. This simulation incorporates both aerodynamic and gravitational factors to precisely replicate the missile's dynamic behavior throughout its flight. To achieve this, key constants and initial conditions, as detailed in Table I, are carefully defined and utilized. The initial conditions are fundamental to defining the starting state of the missile and establishing the foundation for an accurate trajectory simulation. Along with the parameters outlined in Table I, the glide path angle is initialized at -5 degrees, indicating a shallow descent trajectory relative to the horizontal plane. This angle is essential for determining the interplay between the vertical and horizontal components of the missile's motion during its glide phase [8]. The heading angle is set at 0 radians, ensuring that the missile's motion begins in a predefined direction within the horizontal plane.

The missile's initial position in three-dimensional space is specified by the coordinates x=0, y=0, and z=h0. This setup provides a well-defined starting point for solving the equations of motion and modeling the missile's trajectory with precision.

TABLE I.    INITIAL CONDITIONS AND CONSTANTS

| Name | Symbol | Value |
|---|---|---|
| Gravitational constant | G | $3.98 \times 10^{14}$ m$^3$/s$^2$ |
| Radius of Earth | R | 6371000 m |
| Air density at sea level | $\rho_0$ | 1.225 kg/m$^3$ |
| Decay constant | k | $1.41 \times 10^{-4}$ |
| Cross-sectional area of the missile | A | 0.88 m$^2$ |
| Mass of the missile | m | 907 kg |
| Drag coefficient | $C_d$ | 0.5 |
| Lift coefficient | $C_l$ | 0.7 |
| Simulation timestep | $\Delta t$ | 0.1 s |
| total simulation time | $\Delta t_{total}$ | 300 s |
| Initial velocity of the missile | $v_0$ | 5100 m/s |
| Initial altitude of the missile | $h_0$ | 80000 m |

By carefully selecting these initial conditions, the simulation accurately reflects the dynamic environment in which the missile operates, accounting for the influences of gravitational, drag, and lift forces. Any alterations to these initial parameters would significantly affect the resulting trajectory, emphasizing their critical importance in the simulation process. This careful parameterization ensures the reliability and robustness of the trajectory predictions [9].

The motion of a hypersonic missile is influenced by several key forces that govern its trajectory, stability, and overall performance. Understanding these forces is essential for accurately simulating and predicting the missile's behavior during flight. One of the key forces affecting the rocket's motion is gravitational force [10]. This force pulls the rocket toward the ground and varies with the rocket's altitude. The acceleration due to gravity (g) decreases as altitude increases and can be calculated using (1):

$$g = \frac{G}{(R+h)^2} \qquad (1)$$

Here, h represents the missile's altitude. Gravitational force plays a crucial role in influencing the missile's vertical motion and serves as a fundamental factor in determining its overall trajectory path.

The drag force acts in opposition to the missile's motion as it travels through the atmosphere. This force depends on the missile's velocity, cross-sectional area, atmospheric density, and drag coefficient [11]. It is determined using (2):

$$F_d = \frac{1}{2} C_d \rho v^2 A \qquad (2)$$

where $\rho$ is the atmospheric density and v is the velocity of the missile.

Lift force acts perpendicular to the missile's direction of motion and is generated by its aerodynamic design. It provides stability and maneuverability, enabling the missile to adjust its flight path [12]. The lift force is determined using (3):

$$F_l = \frac{1}{2} C_l \rho v^2 A \qquad (3)$$

Atmospheric density is a critical factor in calculating the magnitude of both drag and lift forces. It decreases exponentially with increasing altitude, as described by (4):

$$\rho = \rho_0 e^{-kh} \qquad (4)$$

The interplay of these forces determines the missile's trajectory and stability during flight. Gravitational force drives the descent, while drag slows the missile down, and lift provides directional control. By accurately modeling these forces, it is possible to predict the missile's path, optimize its performance, and design effective countermeasures for interception. A detailed understanding of these forces is therefore essential for improving missile accuracy, stability, and survivability in dynamic environments.

The trajectory of a missile is determined by solving the equations of motion, which describe how the missile moves under the influence of various forces. These equations take into account the combined effects of gravitational pull, aerodynamic forces such as drag and lift, and any additional forces generated by the missile's propulsion system or maneuvering mechanisms. By integrating these equations, the missile's position, velocity, and orientation can be calculated over time, allowing for the prediction of its path through three-dimensional space [13].

The integration process involves breaking the missile's motion into small, discrete time steps. At each step, the forces acting on the missile are recalculated based on its current position, velocity, and orientation. These forces are then used to update the missile's velocity and trajectory for the next time step. This iterative process continues until the missile completes its flight or reaches its target. This approach enables the accurate simulation of a missile's motion, taking into account the influence of changing atmospheric conditions, varying altitudes, and dynamic forces. The results provide a detailed representation of the missile's trajectory, which is essential for understanding its behavior, optimizing its design, and developing strategies for effective defense against such weapons. By using numerical integration, it is possible to model even the most complex trajectories that would be difficult to solve analytically, making it a powerful tool in missile trajectory analysis and prediction.

The rate of change of velocity, as expressed in (5), describes how the missile's velocity evolves over time due to the combined effects of various forces acting on it. This includes the deceleration caused by aerodynamic drag, the downward pull of gravitational force, and the influence of the missile's trajectory angle. Aerodynamic drag, which opposes the missile's motion, depends on factors such as velocity, atmospheric density, and the missile's aerodynamic properties [14]. Gravitational force, which varies with altitude, contributes to the vertical component of velocity change, particularly during the descent phase. The trajectory angle further modulates these forces, determining the balance between vertical and horizontal motion. By integrating this rate of change over time, the missile's velocity can be accurately predicted at each point in its trajectory, providing critical insights into its flight dynamics and overall performance.

$$\dot{v} = \frac{-F_d}{m} - g \sin \theta \qquad (5)$$

where θ is the glide path angle.

The rate of change of the path angle, as represented in (6), quantifies how the missile's flight path angle evolves over time due to the interplay of aerodynamic and gravitational forces. This change is influenced by lift forces, which act perpendicular to the missile's motion, and gravitational components, which vary with altitude and trajectory orientation. The balance between these forces determines whether the missile ascends, descends, or maintains a steady glide. By accurately modeling the rate of change of the path angle, it is possible to predict adjustments in the missile's trajectory, ensuring a precise understanding of its flight behavior and enabling effective trajectory optimization or interception strategies.

$$\dot{\theta} = \frac{F_l}{mv} - \frac{g \cos \theta}{v} \qquad (6)$$

Equation (7) describes the rate of change of the missile's position in three-dimensional space, providing a detailed mathematical framework for modeling its motion along the x, y, and z axes. These equations are expressed as:

$$\dot{x} = v \cos \theta \cos \Phi$$
$$\dot{y} = v \cos \theta \sin \Phi \qquad (7)$$
$$\dot{z} = v \sin \theta$$

$\Phi$ is the heading angle, which defines the direction of motion in the horizontal plane. These equations illustrate how the velocity and orientation of the missile influence its displacement in each spatial dimension. The rate of change along the x-axis ($\dot{x}$) and y-axis ($\dot{y}$) depends on the horizontal component of the velocity, modulated by the flight path and heading angles. In contrast, the rate of change along the z-axis ($\dot{z}$) is governed by the vertical component of the velocity, dictated by the flight path angle.

The trajectory data, including time, x-position, y-position, and z-altitude, is organized into a structured dataset. This dataset is subsequently used for training and evaluating the predictive model. By integrating these equations over time, the missile's trajectory can be reconstructed in three-dimensional space, offering valuable insights into its flight dynamics. This detailed modeling is crucial for tracking the missile's motion, optimizing its design, and enhancing strategies for interception or countermeasures. Equation (7) thus serves as a fundamental tool for accurately predicting the spatial behavior of a missile during its flight.

### B. Missile trajectory prediction model

Artificial neural networks have gained widespread recognition in the field of artificial intelligence due to their ability to perform iterative processes of data analysis, information processing, and pattern recognition. These capabilities make them highly effective tools for prediction tasks, especially in complex and dynamic environments [15].

In this research, a hybrid approach combining CNN, LSTM networks, and GRU has been employed to predict the trajectory of a hypersonic missile. CNNs, LSTMs, and GRUs contribute distinct strengths that make the hybrid model robust for trajectory prediction. CNNs process trajectory data by applying convolutional filters that extract spatial features, identifying localized patterns such as abrupt changes in direction or repetitive motion trends caused by specific maneuvers. This hierarchical approach allows CNNs to capture both basic and complex spatial relationships in the data [13]. LSTMs use gating mechanisms to regulate information flow, enabling them to model sequential dependencies and retain long-term temporal relationships, such as changes in altitude and velocity over time [8]. GRUs, with a simplified architecture, efficiently handle long-term dependencies while reducing computational overhead, making them particularly effective for extended sequences. By integrating spatial and temporal feature extraction, this architecture ensures precise modeling of trajectory dynamics and addresses the challenges posed by predicting the complex flight paths of hypersonic missiles [10].

The implementation of these neural networks has been carried out using the TensorFlow and Keras frameworks, which provide robust tools for building, training, and optimizing deep learning models. This combination ensures that the proposed model is not only effective but also computationally efficient in predicting the complex trajectory dynamics of hypersonic missiles.

The model begins with an input layer that accepts data in the form of sequences, where each sequence includes positional information (x, y, and z) over a defined number of time steps. This input structure allows the model to process both spatial and temporal aspects of the missile's trajectory effectively.

The CNN component extracts spatial features from the input data. Using a 1D convolutional layer with 64 filters and a kernel size of 3, the CNN identifies localized patterns in the trajectory, such as abrupt changes or trends. A ReLU activation function introduces non-linearity, enhancing the model's ability to capture complex relationships. The output of the CNN layer is passed through a dropout layer with a 0.3 dropout rate to reduce overfitting and then flattened into a one-dimensional feature vector for further processing.

The LSTM component captures the temporal dependencies in the trajectory data. With 64 units, the LSTM layer processes the input sequence while maintaining the sequential nature of the data by outputting a sequence of values, retaining detailed temporal information at each step. A dropout layer with a rate of 0.3 is applied to prevent overfitting, followed by a flattening step to prepare the output for combination with other components. This makes the LSTM particularly effective for modeling long-term dependencies critical to trajectory prediction. The GRU component also focuses on temporal relationships but offers a computationally efficient alternative to the LSTM. With 64 units, the GRU processes the input sequence and outputs a single value for the sequence. Similar to the other components, a dropout layer with a rate of 0.3 enhances generalization and prevents overfitting. The GRU effectively complements the LSTM in capturing temporal dynamics.

The outputs from the CNN, LSTM, and GRU components are concatenated into a unified feature vector. This combined representation leverages the strengths of each architecture—spatial feature extraction from CNNs and temporal dependency modeling from LSTM and GRU layers. The concatenated features are passed through a dense layer with 128 neurons and a ReLU activation function, allowing the model to learn higher-order relationships. A dropout layer with a rate of 0.3 is applied before the final output layer, which contains a single neuron with a linear activation function to predict a continuous value representing the next position in one of the dimensions (x, y, and z).

Due to the need for increased accuracy in hypersonic missile trajectory prediction, each positional dimension (x, y, and z) is modeled independently. This results in three separate hybrid models, one for each dimension. Training these models independently ensures that each model specializes in learning the dynamics specific to its respective dimension, further improving prediction accuracy. Each model is trained for 50 epochs with a batch size of 16, balancing computational efficiency with convergence requirements. The training process uses trajectory data divided into sequences of a specified length, with each sequence used to predict the next position in its respective dimension. The models are optimized using mean squared error (MSE) [16] as the loss function and the Adam optimizer for efficient convergence. Mean absolute error (MAE) [17] is used as a performance metric to evaluate the models during training and testing. This hybrid architecture combines the CNN's ability to extract spatial features with the LSTM and GRU's strengths in modeling temporal relationships, making it highly effective for predicting the complex and dynamic trajectories of hypersonic missiles. By independently modeling each dimension, the approach ensures improved accuracy in trajectory predictions. The implementation ensures a robust and scalable framework, capable of addressing the complex challenges associated with hypersonic missile trajectory prediction tasks, while maintaining adaptability for future enhancements.

## IV. RESULTS AND DISCUSSION

Hypersonic missiles, with their extraordinary speed, maneuverability, and precision, are capable of inflicting severe damage on strategic targets, including military installations, critical infrastructure, and urban centers. Their ability to evade traditional missile defense systems makes them a significant threat in modern warfare. Unlike conventional ballistic or cruise missiles, hypersonic weapons can operate at extreme velocities while maintaining high levels of aerodynamic control, allowing them to engage targets with unprecedented speed and accuracy.

As a result, nations worldwide are heavily investing in the development and deployment of hypersonic missile systems to gain a strategic advantage and enhance their deterrence capabilities. In addition to offensive applications, the rise of hypersonic technology has triggered a parallel race in counter-hypersonic defense systems, pushing research in advanced radar networks, satellite tracking, directed energy weapons, and AI-driven interception strategies. Major military powers are also exploring dual-use applications, integrating hypersonic capabilities into next-generation aircraft, space-based platforms, and autonomous weapons systems.

Beyond military applications, the development of hypersonic technology has significant implications for space exploration, high-speed aerospace travel, and rapid-response defense systems. The materials and engineering breakthroughs required for hypersonic flight—such as thermal-resistant coatings, advanced propulsion, and computational fluid dynamics modeling—are paving the way for innovations beyond warfare. However, the rapid proliferation of hypersonic missile programs has also raised concerns over arms control, strategic stability, and global security, with policymakers debating the need for international regulations and new defense frameworks to mitigate the risks of an emerging hypersonic arms race.

In modern warfare, the rapid advancement of hypersonic missile technology has created an urgent need for effective countermeasures. Unlike conventional missile threats, hypersonic weapons challenge existing defense systems by operating at extreme velocities and executing complex flight maneuvers. Their ability to cover vast distances in a short time significantly reduces the window for detection and response, making real-time tracking and interception a critical focus for military defense research. Consequently, the development of advanced counter-hypersonic systems has become a top priority for national security and strategic stability.

One of the most promising approaches to addressing this challenge is the use of neural networks, which have proven to be highly effective in analyzing complex, non-linear data patterns. Unlike conventional computational methods that rely on predefined equations and fixed models, neural networks excel in adaptive learning, meaning they can continuously improve their accuracy by analyzing massive amounts of real-time sensor data. This capability is particularly crucial for predicting missile trajectories, as hypersonic weapons can rapidly change their path mid-flight, making traditional interception calculations insufficient.

These AI-driven models not only enhance missile tracking but also assist in interception planning, response coordination, and target prioritization. As hypersonic missile technology advances, the integration of neural network-based defense frameworks is expected to be a key factor in maintaining strategic balance, military deterrence, and national security in the face of emerging high-speed threats.

The objective of this research is to evaluate the effectiveness of a hybrid neural network model, the CNN-LSTM-GRU architecture, in accurately predicting the trajectory of a hypersonic missile. By leveraging the combined strengths of CNNs, LSTMs, and GRUs, this study aims to contribute to the advancement of predictive technologies essential for ensuring global security.

The accurate simulation of a missile's trajectory is a fundamental step in the analysis and prediction of its flight path, serving as the foundation for realistic modeling and precise forecasting. A well-developed simulation framework allows researchers to model the intricate physical dynamics that govern missile motion, enabling a comprehensive study of various aerodynamic, gravitational, and propulsion-related forces. By replicating real-world flight conditions, these simulations provide invaluable insights into how hypersonic missiles behave under different operational scenarios.

In addition to enhancing theoretical understanding, accurate trajectory simulation is essential for identifying key flight patterns, optimizing guidance algorithms, and refining missile defense strategies. By analyzing simulated trajectories, researchers can detect potential vulnerabilities in flight behavior, assess the impact of external factors such as atmospheric turbulence and electronic interference and countermeasures, and explore countermeasure tactics for interception and neutralization. This capability is particularly valuable in hypersonic missile defense, where precise trajectory modeling is required for real-time tracking, interception planning, and strategic response formulation.

Furthermore, trajectory simulations serve as a crucial component in AI-driven predictive models, providing the training data and validation benchmarks needed for developing machine learning-based forecasting systems. By leveraging high-fidelity trajectory simulations, researchers can enhance neural network-based prediction models, ensuring that AI systems are trained with accurate, diverse, and realistic flight scenarios. This integration of physics-based simulations and AI methodologies strengthens the ability to anticipate missile movements with high precision, ultimately contributing to more effective threat mitigation and national security measures. Simulating trajectories provides a reliable framework for studying and mitigating the significant threats posed by hypersonic missiles.

In this study, as shown in Figure 4, the trajectory of a hypersonic missile was simulated to provide a basis for further analysis and to support the development of advanced predictive modeling techniques.

Figure 4 illustrates the realistic trajectory of a hypersonic missile in 3D space, capturing the defining characteristics that govern the motion of such advanced weapons. The plotted trajectory highlights the distinct flight phases of the missile, showcasing its ability to maintain extreme speeds, perform evasive maneuvers, and reach its target with high precision. The flight path follows a non-ballistic, adaptive course, reflecting the advanced guidance and control systems employed in modern hypersonic weapons.

The initial phase of the trajectory begins at high altitude, representing the missile's launch and acceleration stage. This phase is crucial for enabling the missile to reach hypersonic speeds, typically exceeding Mach 5, by leveraging a boost-phase propulsion system, such as a rocket booster or scramjet engine. During this stage, the missile experiences rapid acceleration as it ascends into the upper atmosphere, where air resistance is minimal, allowing it to achieve optimal velocity while reducing thermal stress and fuel consumption.

Following the initial ascent, the trajectory exhibits a sharp descent, marking the transition into denser atmospheric layers. This phase is a key characteristic of HGVs and hypersonic cruise missiles (HCMs), allowing them to re-enter the lower atmosphere and take advantage of aerodynamic lift and drag forces. The interaction with denser air enhances the missile's ability to maneuver dynamically, making its path less predictable and significantly complicating interception by conventional missile defense systems.

The midsection of the trajectory features an undulating motion, demonstrating the missile's ability to execute high-speed maneuvers. These oscillations arise from a complex interplay of aerodynamic forces, gravity, and lift, allowing the missile to continuously adjust its path while maintaining flight stability.

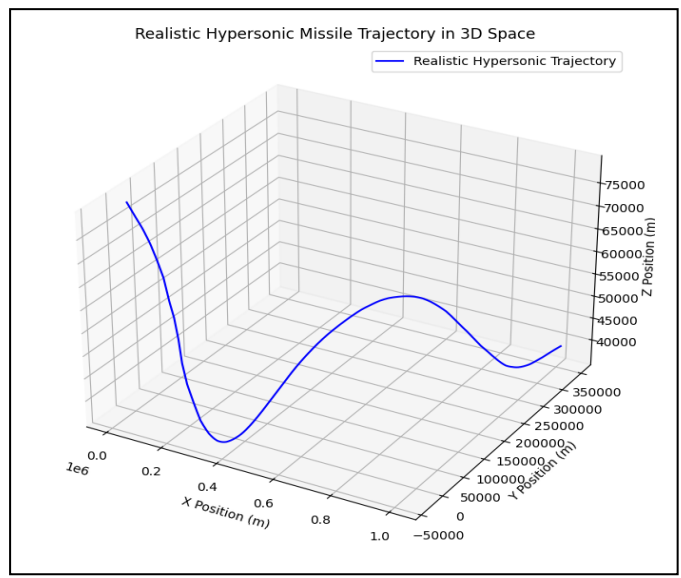

Fig. 4. Hypersonic missile trajectory

Unlike traditional ballistic missiles, which follow a predictable parabolic arc, hypersonic missiles use adaptive flight control to modify their course mid-flight, enhancing their ability to evade detection and interception. These dynamic adjustments are particularly effective in countering ground-based missile defense systems, radar tracking, and anti-aircraft measures.

As the missile progresses toward its terminal phase, the trajectory gradually stabilizes, reflecting the transition to precision-guided targeting. This phase is crucial for ensuring that the missile remains on (course to strike its intended target with high accuracy. The guidance system, which may integrate inertial navigation, satellite-based corrections, and onboard AI-driven adjustments, ensures that the missile maintains a precisely controlled descent. At this stage, final trajectory corrections are made to compensate for any external disturbances, ensuring that the missile delivers its payload effectively.

The overall trajectory observed in Figure 4 encapsulates the core principles of hypersonic missile flight, including rapid initial acceleration, controlled descent, evasive maneuvers, and precise targeting. The ability to combine speed, agility, and unpredictability is what makes hypersonic missiles exceptionally challenging to intercept and positions them as a revolutionary advancement in modern warfare. The complexity of their flight paths underscores the importance of advanced computational models and AI-driven predictive technologies in tracking, predicting, and countering hypersonic threats, reinforcing the strategic necessity of continuous research and development in this field.

To ensure reliable and accurate predictions on new and unseen data, proper division of the dataset into training and test sets is a critical step in the modeling process. The training set plays a fundamental role in building and fine-tuning the model by allowing it to learn underlying patterns and relationships in the data. Conversely, the test set serves as an independent benchmark to evaluate the model's performance on unseen data, providing a realistic measure of its generalization ability.

Without proper evaluation, models that are overfitted to the training data may perform well on familiar datasets but fail to generalize effectively to new scenarios. By testing the model on a separate dataset, researchers can assess its accuracy, identify weaknesses, and implement necessary adjustments or improvements to enhance its robustness.

In this study, the dataset was partitioned into training and test sets, with 80% of the trajectory data allocated for training the model and the remaining 20% reserved for testing purposes. This approach ensures that the model's performance is evaluated comprehensively and under conditions that closely mimic real-world applications.

The performance of the CNN-LSTM-GRU hybrid neural network in predicting the trajectory of a hypersonic missile has been evaluated using standard error evaluation criteria. The results of this evaluation are presented in Table II, which provides a quantitative measure of the model's predictive accuracy.

Additionally, Figure 5 illustrates the predicted trajectory of the hypersonic missile, offering a visual representation of the model's output.

TABLE II. FORECAST ACCURACY MEASURES FOR MODELS

| Accuracy Measure | CNN-LSTM-GRU |
|---|---|
| Root Mean Squared Error (RMSE) | 3330.46 |
| MAE | 3110.06 |
| Mean Absolute Percentage Error (MAPE) | 1.83 |

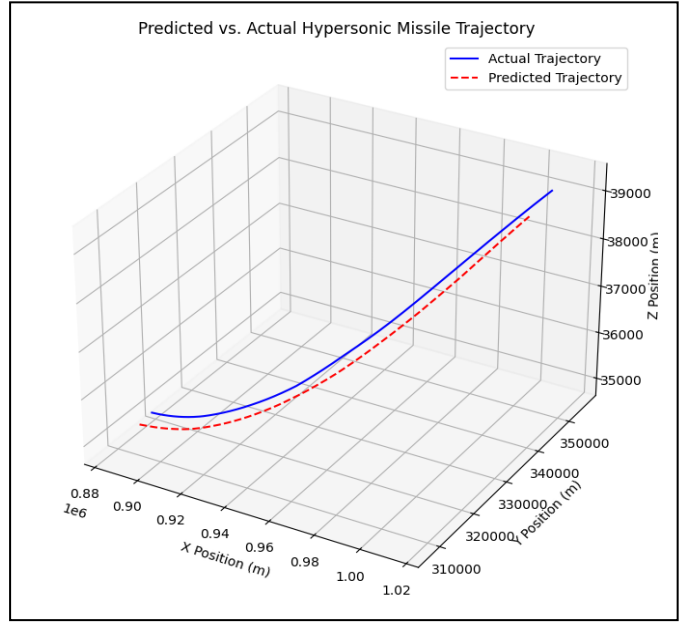

Fig. 5. Hypersonic missile trajectory prediction

These results serve as a foundation for further analysis and validation of the model's effectiveness in handling complex trajectory prediction tasks. The model achieved an RMSE of 3330.46, an MAE of 3110.06, and a MAPE of 1.83%, demonstrating its high accuracy and minimal deviation in predicting complex hypersonic missile trajectories. These results indicate the model's ability to effectively capture the highly dynamic, non-linear flight behavior of hypersonic missiles, which are influenced by rapid velocity changes, aerodynamic forces, and external environmental factors. The low error margins validate the model's effectiveness in handling trajectory fluctuations and maintaining precision across different flight phases.

Figure 5 provides a visual comparison between the predicted and actual missile trajectories, illustrating how well the model approximates the missile's flight path. The blue curve represents the actual trajectory derived from ground truth flight data, while the red dashed curve corresponds to the predicted trajectory generated by the CNN-LSTM-GRU model. The close alignment between these two curves, with only minor deviations, highlights the model's ability to generalize well across different test scenarios. The limited gap between the predicted and actual paths confirms that the model is capable of precisely forecasting trajectory variations, even in the presence of nonlinear aerodynamic effects and unpredictable disturbances.

The combination of quantitative results in Table II and the graphical representation in Figure 5 reinforces the robustness and reliability of the proposed model in handling intricate trajectory prediction tasks. The low RMSE, MAE, and MAPE values, coupled with the strong visual correlation between predicted and actual flight paths, underscore the model's effectiveness in real-world applications, particularly in missile trajectory forecasting, guidance systems, and defense strategy formulation. The ability to achieve such high accuracy is crucial in military and aerospace applications, where precise trajectory prediction is essential for early warning systems, interception planning, and defense response coordination.

Furthermore, the model's performance suggests that deep learning-based hybrid architectures, such as the CNN-LSTM-GRU, offer a powerful and scalable solution for trajectory analysis in high-speed aerial systems. As hypersonic technology continues to advance, integrating such predictive models into real-time tracking frameworks and defense systems can significantly enhance situational awareness and decision-making capabilities in missile defense operations.

## V. CONCLUSION AND FUTURE WORK

Hypersonic missiles represent a significant technological leap in modern warfare, characterized by their extraordinary speeds and agility, which challenge conventional defense systems. The ability to accurately predict their trajectories is crucial for designing effective countermeasures and safeguarding against potential threats. This research presented a robust approach to trajectory prediction by employing a hybrid neural network architecture that combines CNNs, LSTM networks, and GRUs. The proposed model achieved impressive predictive performance, as evidenced by a RMSE of 3330.46, MAE of 3110.06, and a MAPE of 1.83%. These results underscore the efficacy of the hybrid model in addressing the complexities of hypersonic missile dynamics.

In future work, several avenues can be explored to enhance and expand this research. Incorporating real-world data from advanced sensors and integrating domain-specific physics into the modeling process could improve the accuracy and reliability of predictions. Additionally, the application of transfer learning techniques may enable the model to adapt to various missile types and operating environments more effectively. Further studies could also explore the integration of this predictive framework into real-time defense systems for immediate application in missile interception and countermeasures. By pursuing these directions, the predictive capabilities of defense systems can continue to evolve, providing a robust response to the challenges posed by hypersonic threats.

ACKNOWLEDGMENT

During the preparation of this work, the author used ChatGPT to edit the text and improve the English language. After utilizing this tool, the author reviewed and modified the content as necessary and takes full responsibility for the final version of the publication.